\documentclass[prb,amsmath,superscriptaddress,12pt,tightenlines]{revtex4}

\usepackage{bm}

\usepackage{color}

\usepackage{graphicx} 
\usepackage{multirow}
\usepackage{amsmath}

\usepackage{amsfonts}
\usepackage[version=3]{mhchem}

\usepackage{dcolumn}
\usepackage{bm}

\usepackage{dcolumn}

\newcommand{\dif}{\mathrm{d}}

\newcommand{\muexBa}{\mu^{\mathrm{(ex)}}_{\mathrm{Ba}^{2+}}}
\newcommand{\muexSr}{\mu^{\mathrm{(ex)}}_{\mathrm{Sr}^{2+}}}

\begin{document}
\setcounter{page}{1} 
\title{Strontium and Barium in Aqueous Solution and a Potassium Channel Binding Site}
\author{Mangesh I. Chaudhari}
\email{michaud@sandia.gov}
\affiliation{Center for Biological and Engineering Sciences, Sandia National
Laboratories, Albuquerque, NM 87185, USA}
\author{Susan B. Rempe}
\email{slrempe@sandia.gov}
\affiliation{Center for Biological and Engineering Sciences, Sandia National
Laboratories, Albuquerque, NM 87185, USA}

\date{\today}

\begin{abstract}%
{Ion hydration structure and free energy establish criteria
for understanding selective ion binding in potassium (K$^+$) ion channels, and may be significant to understanding blocking mechanisms as well. 
Recently, we investigated the hydration properties of Ba$^{2+}$, the most potent blocker of 
K$^+$ channels among the simple metal ions.  Here, we use a similar method of combining 
$ab$ $initio$ molecular dynamics simulations, statistical mechanical theory, 
and electronic structure calculations to probe the fundamental
hydration properties of Sr$^{2+}$, which does not block bacterial K$^+$ channels. 
The radial distribution of water around Sr$^{2+}$ suggests a stable 8-fold geometry in the local
hydration environment, similar to Ba$^{2+}$.  
While the predicted hydration free energy  of $-$331.8 kcal/mol is comparable with the experimental result of $-$334 kcal/mol, 
the value is significantly more favorable than the $-305$~kcal/mol hydration free energy of Ba$^{2+}$.  When placed in the innermost
K$^+$ channel blocking site,  the solvation free energies and lowest energy structures of both Sr$^{2+}$ and Ba$^{2+}$ are nearly
unchanged compared with their respective hydration properties.  That result suggests that differences in blocking behavior arise due to 
free energies associated with exchange of water ligands for channel ligands instead of free energies of transfer from water to the binding site.
}

\end{abstract}
\maketitle

\section*{Introduction}

Molecules that bind to potassium 
(K$^+$) channels
and block K$^+$ ion permeation across cellular membranes
can have detrimental or beneficial effects. 
Peptide toxins from several poisonous animals provide an example
of the former.\cite{banerjee,rosenbaum} For the latter case,
drugs that block specific K$^+$
channels hold promise for treating neurological 
disorders,\cite{wulff} autoimmune diseases,\cite{wulff, he:2015} 
and cancers.\cite{pardo} 
Alkaline earth metals also block K$^+$ channels. 
As the simplest of the blocking molecules, 
divalent ions provide a favorable case for understanding the current 
issue of channel block.  Two prominent ions 
used in K$^+$ channel blocking studies
are strontium (Sr$^{2+}$) and barium (Ba$^{2+}$). 
Comparable ion size and additional charge compared to the permeant K$^+$ ion has been proposed as an explanation for  block by Ba$^{2+}$.\cite{Jiang:2000,Piasta:2011bu}  That explanation naturally focuses attention on ion solvation by channel binding sites and free energies of ion transfer between water and those binding sites.

\begin{figure}
\includegraphics[width=3.0in]{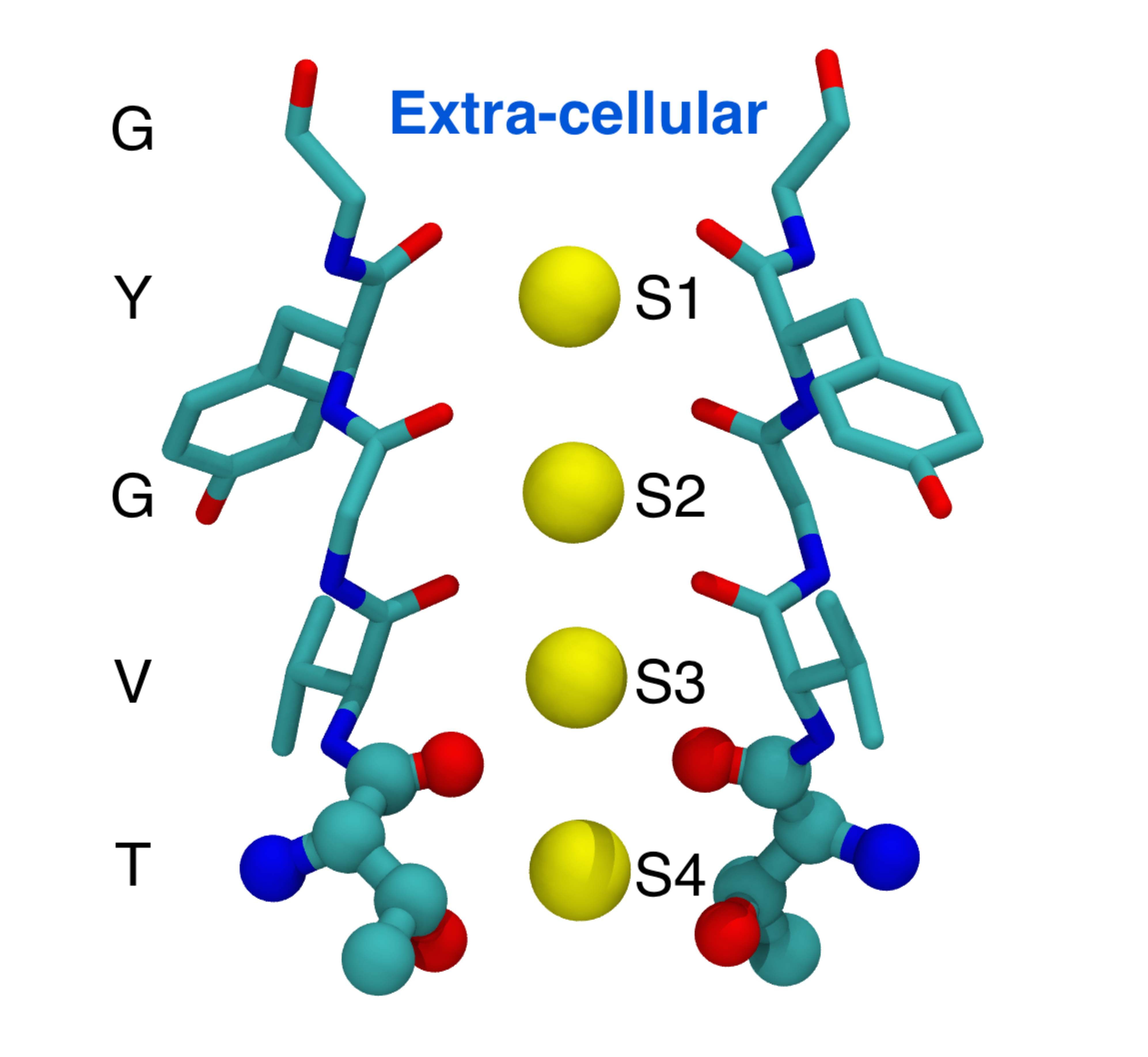}
\caption{Sr$^{2+}$ and Ba$^{2+}$ ions bind to the innermost site (S4)
in the conserved selectivity filter of mammalian potassium ion channels, where four threonine (T)
residues with eight oxygens (red) surround an ion.\cite{{Sugihara:1998vp,Soh:2002ga,Jiang:2000,lockless,ye,Jiang:2014}}  Unlike Ba$^{2+}$, Sr$^{2+}$ does not bind to  S4  in bacterial potassium ion channels. For clarity, only two of the four monomers from the crystallized bacterial KcsA channel (occupied by K$^+$) are shown.\cite{Zhou:2001vo}} 
\label{fig:Kchannel}
\end{figure}

Despite close similarity of 0.16~{\AA} in crystal radius and
identical +2 charges,\cite{shannon} 
 Ba$^{2+}$ and Sr$^{2+}$ ions exhibit different blocking 
behaviors in K$^+$ channels. As the most potent of the metal ion blockers,  Ba$^{2+}$ binds
preferentially to the innermost (S4) ion binding site
in bacterial and eukaryotic K$^+$ channels.\cite{Armstrong:1982,Jiang:2000}  Ba$^{2+}$ binding induces non-conducting events 10-1,000-fold longer-lived than  spontaneous channel closings, which allowed analysis of binding equilibria for K$^+$ and Na$^+$ ions.\cite{Piasta:2011bu}
X-ray
crystallographic studies of bacterial channels
report that Ba$^{2+}$ fully dehydrates upon transfer from
aqueous solution and adopts
the same octa-ligated conformation in the S4 binding site as K$^+$
(FIG.~\ref{fig:Kchannel}).\cite{Jiang:2000,lockless,ye,Jiang:2014} That similarity in solvation structure supports the proposed mechanism of divalent ion block.

In contrast, the slightly smaller
Sr$^{2+}$ ion binds
less favorably to the S4 
blocking sites of mammalian K$^+$ channels \cite{Sugihara:1998vp,Soh:2002ga} 
and does not block 
bacterial K$^+$ channels.\cite{Piasta:2011bu}
The difference in blocking behavior
between Sr$^{2+}$ and Ba$^{2+}$ lacks an explanation and also undermines the traditional explanation for block.  Could that difference be attributable to differences in hydration properties of these ions?  Or could the difference instead be related 
to different solvation properties
within the conserved S4 binding site of K$^+$ channels?

Toward answering these questions,  we previously  reported Ba$^{2+}$ hydration structure and free energy.\cite{Chaudhari:2014wb} 
Here, we report a similar study of Sr$^{2+}$ hydration.  In addition, we carry out a direct comparison of  Sr$^{2+}$ and Ba$^{2+}$ solvation properties  at the innermost K$^+$ channel blocking site, S4.
To facilitate the comparison, we adapt the quasi-chemical free energy theory to describe solvation by the S4 binding site.  In particular, we compute the absolute ion solvation free energies in water and the channel binding site to validate our approach and gain more insight over relative values computed more commonly.  
 Our results emphasize close similarities between the two divalent ions in terms of solvation structure and free energy in the two distinct environments, as expected for  ions of similar size and equivalent charge.  That leads to the conclusion that  differences in blocking behavior of Ba$^{2+}$ and Sr$^{2+}$ arise due to differences in dehydration and rehydration free energies, 
associated with exchange of water ligands for channel ligands, instead of free energies of transfer from water to the binding site.
 
\section*{Results and Discussion}

\subsection{Sr$^{2+}$ Hydration Structure}
\begin{figure}
\centering	
\includegraphics[width=3.0in]{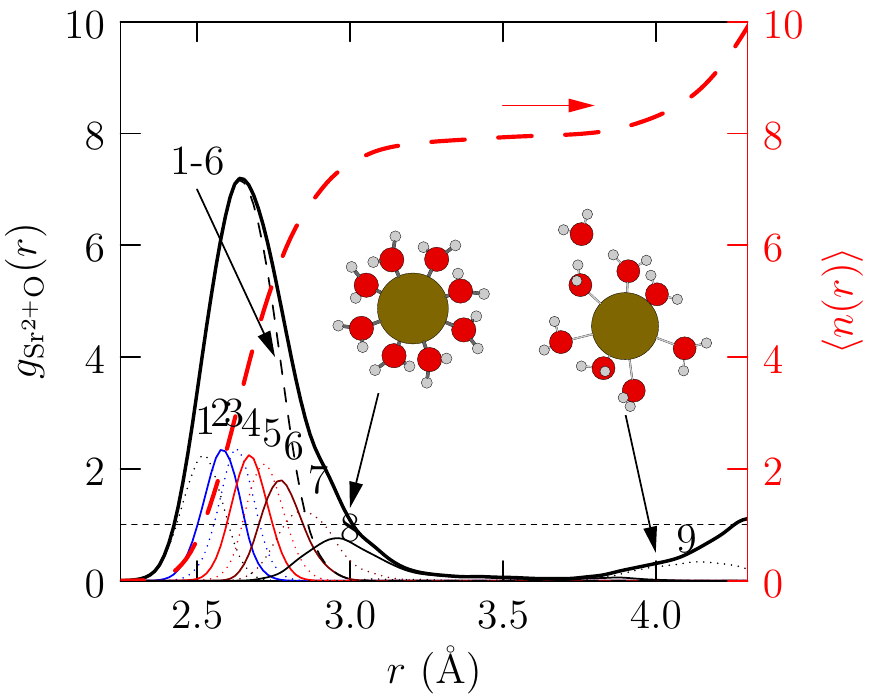}
\caption{The ion-oxygen
radial distribution function and near neighbor distributions for a single Sr$^{2+}$ ion solvated by 64 waters. 
The picture illustrates
an observed conformation of Sr$^{2+}$ ion (brown) with $n$=8 and 9 waters (red oxygens, silver hydrogens). 
Energy optimization for those clusters 
confirms that 
the 9$^{th}$ water occupies the second solvation shell, meaning the cluster is in a
mixed-shell $n$=8+1 conformation rather than the prescribed inner-shell conformation for evaluation of Eq.~1. 
$\left<n(r)\right>$=$4\pi\rho_{\mathrm{H_2O}}{\int_{0}^{r}}g_{\mathrm{Sr^{2+}O}}(x)x^2\dif{x}$ represents 
the running coordination number (red dashed line).  The plateau in $\left<n(r)\right>$ reaffirms that
eight near water neighbors stably occupy the first hydration shell of Sr$^{2+}$(aq) and define a completely filled inner-shell structure.} \label{fig:gr}
\end{figure}

To probe strontium hydration structure, we carried out $ab~initio$ molecular dynamics (AIMD) simulations and calculated radial distribution functions (RDF) between Sr$^{2+}$ and water oxygens (O) (FIG.~\ref{fig:gr}). The first shell is highly structured and well-defined, with eight (8) near-neighbor water molecules forming an inner hydration structure.  A plateau on the running coordination number further confirms this tightly coordinated structure. The first minimum is spread over a broad range (3.4-3.8 \AA), which is common for divalent \cite{Chaudhari:2017gs} and small monovalent ions.\cite{Rogers,Mason:2015} 

Compared to Ba$^{2+}$, Sr$^{2+}$ has a more narrow and well-defined first hydration shell.\cite{Chaudhari:2014wb}  Also,  the position of the first peak (2.63~\AA) is slightly closer for Sr$^{2+}$ than for hydrated Ba$^{2+}$ (2.80~\AA). 
Despite those small differences, eight (8) waters occupy the first hydration shells in both cases.  That number differs from the first hydration shell of K$^+$, which contains a total of 6 waters \cite{Soper:2006,Varma:2006} with only 4 tightly coordinated near neighbors that interact directly with the ion.\cite{Rempe:K} 

These simulation results compare well with experimental results for Sr$^{2+}$, 
\cite{Ramos:2003wd,Pfund:2002bf,ohtaki} as was also the case for Ba$^{2+}$ in earlier work.\cite{Chaudhari:2014wb} 
A slightly lower first shell occupancy of 7.5 obtained by a previous AIMD study \cite{Harris:2003jq} may be attributable to limited sampling from shorter simulation times (4~ps) compared with the current work (30~ps). The differences in RDF are consistent with the smaller ionic radius of Sr$^{2+}$ compared to Ba$^{2+}$ , which pulls water molecules closer to the ion with higher charge density.

\subsection{Theory for Hydration Free Energy}
To evaluate hydration free energy, we applied quasi-chemical theory (QCT),~\cite{redbook,Asthagiri:2010,Rogers} a complete statistical mechanical framework well-known for calculating such values.\cite{Sabo:2013gs,Stevens:2016,Chaudhari:2017}
This 
theory separates inner-shell interactions from outer-shell in a ``divide~and~conquer" approach to obtain the solute excess chemical potential,   
\begin{equation}
	\mu^{\mathrm{(ex)}}_{\mathrm{Sr}^{2+}} = -kT\mathrm{ln}K^{(0)}_{m*n}\rho^{m*n}_{\mathrm{H_2O}}+kT\mathrm{ln}p_{\mathrm{Sr}^{2+}}(m*n) 
\\  +(\mu^{\mathrm{(ex)}}_{\mathrm{Sr(H_2O)}_{m*n}^{2+}}-n\mu^{\mathrm{(ex)}}_{\mathrm{H_2O}}).
\label{eq:1}\end{equation}
Detailed explanations of these terms and derivation of Eq.~\ref{eq:1} are documented elsewhere.\cite{Beck:2006wp,redbook} In anticipation of ion solvation by ligands with multi-dentate modes of ion ligation (as in FIG.~1), we include a denticity factor $m$ not given in earlier works.

Briefly, the first term contains the equilibrium constant $K_{m*n}^{(0)}$ for association of $n$ ligands with the ion to form an inner-shell cluster in an ideal gas, denoted by the superscript (0).  An inner shell exists when ligands (e.g., waters) interact directly with the ion. If the ligand is multi-dentate, then $m$ specifies the number of functional groups that coordinate with the ion ($m$=1 for water).  
This association free energy term is a well-defined computational target that naturally takes into account full quantum chemical effects relevant to the chemical strength interactions between ions and ligands.  Here we adopt standard electronic structure software to evaluate this term and apply well-studied harmonic approximations \cite{Rempe:1998,Rogers:2011} based on mechanically stable cluster structures. The harmonic approximation takes into account small atomic displacements from structures that represent minima on potential energy surfaces.

The  density term, $\rho^{m*n}_\mathrm{H_2O}$, accounts for the availability of molecules ($n$ waters with $m$=1 functional groups, in this case) to act as ligands of the ion. 

The second term gives the thermal probability, $p_{\mathrm{Sr}^{2+}}(m*n)$, of observing an $n$-fold cluster in solution that ligates the ion with $m$ functional groups.  This probability can be estimated from AIMD simulation data. If only a single complex forms in solution, then the probability of observing that complex is 1 and the free energy contribution is zero.  That situation can occur in strongly bound complexes formed from ligands interacting with multivalent ions such as Sr$^{2+}$ or Ba$^{2+}$.  Alternatively, interactions between the cluster and surrounding environment may constrain the structure of the binding site.  Structural constraints occur in many molecules, including ion channels.\cite{Varma:2007bj,Varma:2011ho}  While Eq.~1 can be written to account for any general structural constraint,\cite{Rogers} here we only consider constraints on ligation number ($m*n$).  Since the value of the second term is negligible compared with the other terms, we neglect it here for computation of Sr$^{2+}$ hydration free energy.  

The third term accounts for the outer-shell solvation environment (e.g., here, liquid water) by taking the difference in excess chemical potentials between a cluster Sr(H$_2$O)$_{m*n}^{2+}$ and $n$ ligands. The clusters are treated as flexible molecular constituents of the solution.  Evaluation of this term typically uses well-studied dielectric models, such as the polarizable continuum model (PCM) applied here. \cite{Tomasi:2005tc} That implicit solvation model is an important approximation, but one that is balanced by the differences taken in the overall combination in Eq.~1.    

\begin{figure}
\includegraphics[width=3.0in]{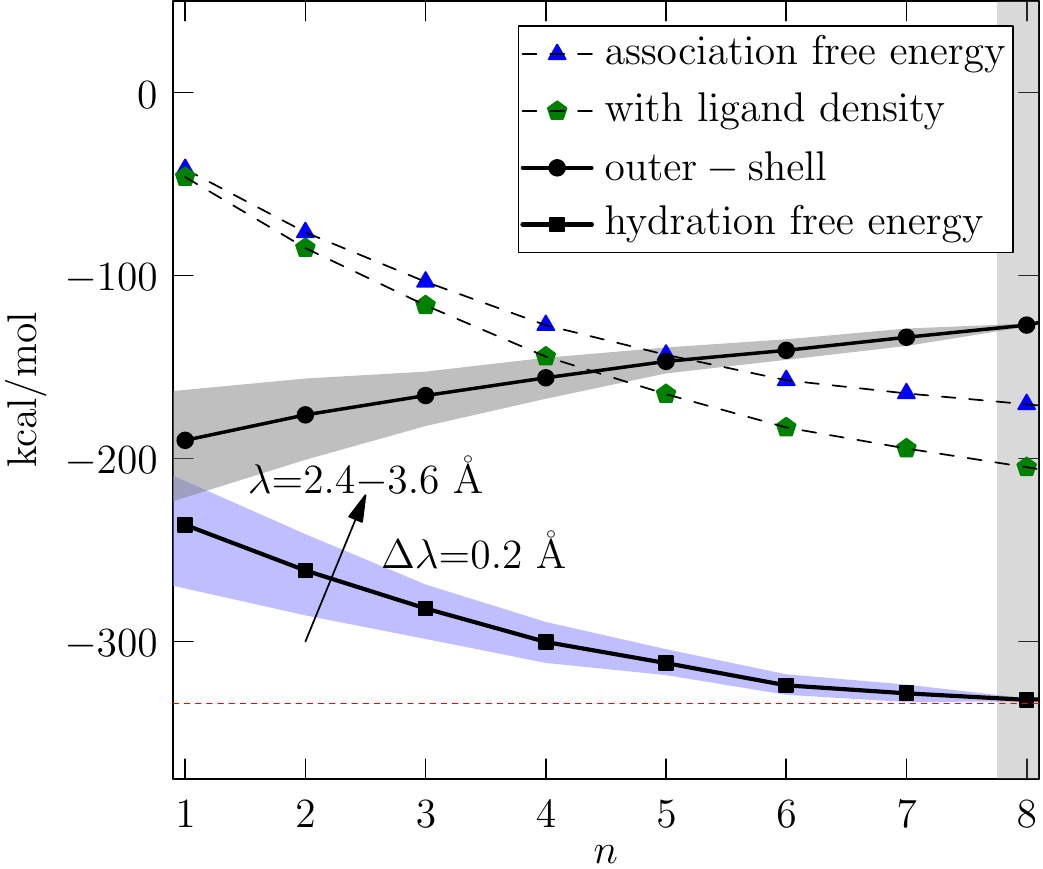}
\caption{Hydration free energy and contributions of terms defined in Eq.~\ref{eq:1} for different inner-shell occupancies, $n$. Horizontal shaded regions cover contributions for each $\lambda$ value (refer to FIG. S1 for entire data set), black symbols are the mean outer-shell contributions and mean hydration free energies calculated for all $\lambda$ values. The
 red dashed line represents an experimental value of $\muexSr$=-333.62 kcal/mol.\cite{Marcus:1994ci} The best prediction comes from treating the completely filled inner-shell occupancy at $n$=8 (light gray vertical shading).  Then the result is independent of inner-shell boundary and agrees with experiment.} 
\label{fig:diff_contri}
\end{figure} 

Selection of an inner-shell occupancy and boundary are essential ingredients of QCT. For this hydration study, we  explore a range of occupancies ($n=$1$-$8) and boundary radii ($\lambda = 2.4$ to $3.6~\mathrm{\AA}$) chosen from features in the radial distribution function, specifically the first hydration shell (FIG.~ \ref{fig:gr}). This elaborated study of the entire first hydration shell has not been done for Sr$^{2+}$. 

\subsection{Sr$^{2+}$ Hydration Free Energy}

Application of Eq.~1  produces a range of ion hydration values (FIG.~3). The variation depends on choice of occupancy and boundary used to define the inner hydration shell.  
Choosing a completely filled inner shell with $n$=8 waters predicts a hydration free energy of $\langle\mu^{\mathrm{ex}}_{\mathrm{Sr}^{2+}}\rangle$ = $-$331.8 kcal/mol that is independent of boundary radii and comparable to the experimental value of $-$333.6 kcal/mol.\cite{Marcus:1994ci} The good agreement arises because a completely filled inner solvation complex reduces errors associated with the outer-shell term (Eq.~1).  Specifically, a filled inner shell  best satisfies the criteria for successful application of an implicit solvent model of the outer-shell solvation environment.\cite{Chaudhari:2014wb}

Decomposition of $\mu^{\mathrm{(ex)}}_{\mathrm{Sr}^{2+}}$ into inner and outer components (FIG.~\ref{fig:diff_contri}) illustrates that all  terms of Eq.~\ref{eq:1}  naturally vary as a function of inner-shell occupancy. Similar to observations of Li$^+$,\cite{Rempe:Li} Na$^+$,\cite{Rempe:Na} K$^+$,\cite{Rempe:K} Rb$^+$,\cite{Sabo:2013gs} Ba$^{2+}$\cite{Chaudhari:2014wb}, and F$^-$\cite{Chaudhari:2017} hydration by QCT analysis, the combined association free energy and ligand density term becomes more favorable as the number of ligands increases.  Simultaneously, the balance of outer-shell contributions (difference between cluster and $n$ ligands) becomes less favorable with increasing $n$.  
The outer-shell contribution varies with inner-shell boundary
until formation of the fully occupied inner hydration shell at $n$=8 (FIG.~3). Then QCT, utilized with the approximations described above, produces an accurate ion hydration free energy that is also independent of inner-shell boundary. 

To summarize, Eq.~1 evaluated earlier with Ba(H$_2$O)$_{1*8}^{2+}$ and presently with Sr(H$_2$O)$_{1*8}^{2+}$ inner-shell clusters coupled to a polarizable continuum model of the surrounding environment produces accurate ion hydration free energies (FIG.~3).


\subsection{Local Ion Hydration vs. Channel Solvation Structure}
Next, we consider ion hydration structure in the context of K$^+$ channels. Earlier works proposed that the octa-coordinated ion binding sites apparent in X-ray studies of K$^+$ channels (FIG.~1) mimic the local hydration structure of 
permeant ions like K$^+$ or Rb$^+$.\cite{Zhou:2001vo}
Studies of ion hydration by \emph {ab initio} molecular dynamics showed that K$^+$ and Rb$^+$ local hydration structures are less than 8-fold coordinated.
\cite{Chaudhari:2014wb}  In other words, the K$^+$ channel binding sites over-coordinate the permeant ions.\cite{Varma:2011ho,Varma:valino}  The binding sites mimic instead the hydration structure of Ba$^{2+}$, a blocking ion. We find a similar result for Sr$^{2+}$ ion. Like Ba$^{2+}$, Sr$^{2+}$ forms stable 8-fold skewed cubic water clusters (FIG.~2). 

From the point of view of drug design, the local ligand architecture of both divalent ions in water mimics binding sites for permeant  ions in K$^+$ channels (FIG.~1).  The similarity in local structure suggests that both divalent ions should fit similarly in potassium ion channel binding sites.  

\subsection{Theory for Solvation by Channel Binding Sites}

To examine ion solvation by K$^+$ channels, we  adapted the  QCT  analysis (Eq.~1) for Ba$^{2+}$ and Sr$^{2+}$ solvation by the S4 binding site (FIG.~1).  As a first task, we constructed the explicit inner-shell solvation structures based on information from the K$^+$ channel crystal structure with the S4 site occupied by Ba$^{2+}$.\cite{lockless}  Specifically, the inner-shell solvation structure consists of four ($n$=4) threonine (THR) amino acid residues that ligate the ion in bidentate mode ($m$=2). Based on that information, we constructed inner-shell clusters Ba(THR)$_{2*4}^{2+}$ and Sr(THR)$_{2*4}^{2+}$.  

To validate these inner-shell structural models, we calculated the root-mean-squared deviation (RMSD) in distance of oxygen atoms between the THR residues of the crystal structure (1K4C) and the optimized Sr$^{2+}$(THR)$_{2*4}$ and Ba$^{2+}$(THR)$_{2*4}$ structures. All oxygen atoms moved less than 1 \AA~after optimization of the electronic energy (TABLE~\ref{table1}). This result confirms that the optimized structures provide a reasonable representation of local structure for K$^+$ ion channels occupied by  Ba$^{2+}$ in the S4 binding site (FIG.~5). 

\begin{table}\centering
\begin{tabular}{|c|c|c|} \hline
Atom&Sr$^{2+}$ (\AA)& Ba$^{2+}$ (\AA)\\ \hline
O1&0.97&0.96\\
O2&0.30&0.16\\
O3&0.98&0.97\\
O4&0.30&0.16\\
O5&0.94&0.95\\
O6&0.55&0.18\\
O7&0.90&0.91\\
O8&0.56&0.22\\
\hline
\end{tabular}
\caption[]
{The RMSD of oxygen atoms in energy-optimized S4 binding sites occupied by the divalent ions studied here,
Sr$^{2+}$(THR)$_{2*4}$ and Ba$^{2+}$(THR)$_{2*4}$, compared with oxygens in the 1K4C crystal structure, K$^{+}$(THR)$_{2*4}$.}\label{table1}
\end{table}

The inner-shell solvation structures represent simplified models of the ion channel.  The study of simplified binding site models, with implicit or explicit inclusion of environmental effects, provides a natural route for learning about the underpinnings of selective ion binding under study here. Simplified models have been used in earlier studies of K$^+$ vs. Na$^+$ selectivity in similar models of interior K$^+$ channel binding sites,\cite{Varma:2007bj} as well as binding sites formed of constrained water molecules.\cite{Rogers:2011} In addition, the same model  was used recently to study Ba$^{2+}$ binding to variants of the S4 binding site.\cite{rossi} Several new features differentiate the current work from those prior studies.  First, we present for the first time the quasi-chemical theoretical framework for ligands with multi-dentate binding modes.  Second, we account for the presence of liquid water in the outer-shell environment adjacent to the S4 binding site. 

In the QCT analysis, the density term describes the availability of ligands to coordinate the ion.  In analogy to the hydration studies, $\rho^{(2*4)}_{\mathrm{THR}}$ may be taken as the density of pure liquid threonine at standard conditions.  Liquid threonine density is the same order of magnitude as liquid water density ($\sim$ 1~g/ml). Given that eight (8) oxygen ligands form a stable complex around both Ba$^{2+}$ in water \cite{Chaudhari:2014wb} and in the S4 binding site,\cite{lockless} the assignment of ligand density appears reasonable.  A precise value for $\rho^{(2*4)}_{\mathrm{THR}}$ is unimportant since the contribution to free energy depends on the natural log of density, $\mathrm{ln}\rho^{(2*4)}_{\mathrm{THR}}$.  

Effects from the surrounding environment are taken into account in the probability and excess chemical potential terms. In the hydration studies, the calculated probability of observing the $n$=8 complex is nearly 1 for both Sr$^{2+}$ and Ba$^{2+}$,\cite{Chaudhari:2014wb} making that term negligible in Eq.~\ref{eq:1}.  Similarly, X-ray crystallography studies report a single structure, the $n$=8 complex around Ba$^{2+}$ in the S4 binding site,\cite{lockless} suggesting that the probability term is also negligible in the channel environment for the 8-fold coordination case.  

The excess chemical potential terms in the hydration problem modeled the aqueous environment as a polarizable continuum with dielectric value $\epsilon$=80.  An aqueous environment borders the S4 binding site on the intracellular side in the form of a cavity with $\sim$20 waters.\cite{Zhou:2001vo}  Toward the extracellular side, a single water  borders the S4 binding site.\cite{Zhou:2001vo} A previous study of Ba$^{2+}$ in the S4 binding site approximated the environment with $\epsilon$=1.  Here instead we account approximately for the neighboring aqueous environment  with $\epsilon$=80.   The resulting ion solvation free energy is relatively insensitive to the precise value of $\epsilon$ for values between 80 and 20. 
Furthermore, errors balance in taking the difference between excess chemical potentials of cluster and ligands. 
  A more realistic model of the environment would treat the aqueous environment on one side of the S4 binding site differently from the other side, but that more complex model is reserved for future work.

The resulting QCT formula for ion (X) solvation by the S4 binding site  describes 8 oxygen atoms from $n$=4 THR residues that ligate the ions in bi-dentate mode ($m$=2),

\begin{equation}
	\mu^{\mathrm{(ex)}}_{\mathrm{X} \mathrm{-S4}} = -kT\mathrm{ln}K^{(0)}_{{2*4}}\rho^{(2*4)}_{\mathrm{THR}}+kT\mathrm{ln}p_{\mathrm{X}}(2*4) 
\\  +(\mu^{\mathrm{(ex)}}_{\mathrm{X(THR)}_{{2*4}}}-4\mu^{\mathrm{(ex)}}_{\mathrm{THR}}).
\label{eq:2}\end{equation}
Since the probability term $p_{\mathrm{X}}(2*4)$ is likely negligible for the 8-fold coordination cases considered here, only the first term, the association free energy, and the last term, the outer-shell free energy, contribute to the ion solvation free energy  in the S4 binding site, $\mu^{\mathrm{(ex)}}_{\mathrm{X} \mathrm{-S4}}$.

\subsection{Ion Transfer Between Water and Channel Binding Sites}
In the context of free energies for Sr$^{2+}$ and Ba$^{2+}$  transfer between water and K$^+$ channel binding sites, 
the slightly smaller Sr$^{2+}$ requires an additional $\sim$30 kcal/mol to take the ion out of the hydration environment (FIG.~4). The additional stability comes mainly from a stronger association free energy for Sr$^{2+}$ in water relative to Ba$^{2+}$, as expected due to the smaller size and higher charge density of Sr$^{2+}$. This additional free energy must be compensated by the channel for both ions to have similar binding affinities. 

Similar to the hydration results,  Sr$^{2+}$ is 30 kcal/mol more favorably solvated in the S4 binding site than Ba$^{2+}$ (FIG.~4).  Again, the additional stability comes from a stronger association free energy for the smaller Sr$^{2+}$ in the S4 binding site.  Similar outer-shell contributions are expected for the two ions in a given environment, as observed (FIG.~4), due to the similarity in cluster size, as well as identical ligand chemistry and number of ligands.   

\begin{figure}
\includegraphics[width=3.0in]{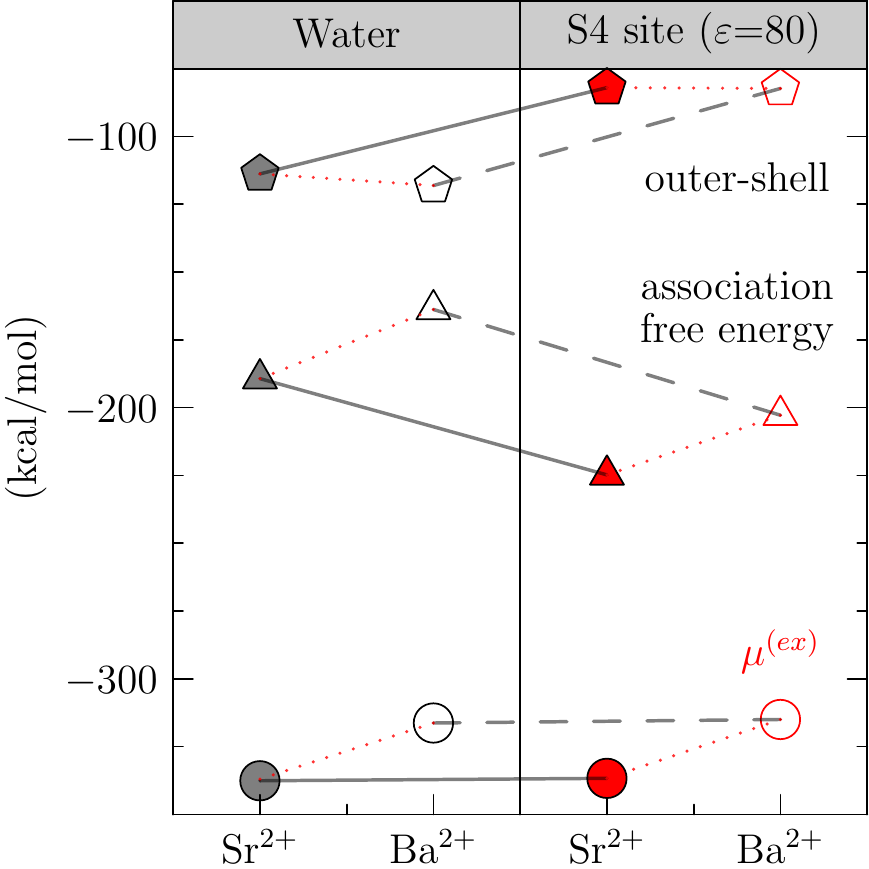}
\caption{Comparison between solvation free energy (Eq.~1) of Sr$^{2+}$ and Ba$^{2+}$ in water (left) and at the S4 
 binding site (right). The transfer free energy is the difference (binding site minus water). The outer shell is modeled as a dielectric of $\epsilon$=80 in both cases because a water-filled cavity lies next to S4. Transfer free energies are $\sim$~0 for both ions.} 
\label{fig:wat_chan_ene_V2_new}
\end{figure}

  Comparing ion solvation in the S4 binding site relative to water ($\mu^{\mathrm{(ex)}}_{\mathrm{X} \mathrm{-S4}}$ versus $\mu^{\mathrm{(ex)}}_{\mathrm{X}}$), 
  the calculated solvation free energies show only minor differences between values 
  of the same ion  (FIG.~ \ref{fig:wat_chan_ene_V2_new}). Interestingly, components of the free energy show substantial differences.  The association free energy for both Sr$^{2+}$ and Ba$^{2+}$ in the S4 binding site of threonine oxygens is 30~kcal/mol more favorable than in water due to the stronger ion-ligand interactions from threonine oxygens.  In contrast, the outer-shell contributions are $\sim$ 30~kcal/mol less favorable in S4 due to the larger size of the ion-ligand clusters in the S4 binding site. These opposing trends lead to a transfer free energy of $\sim$~0 for both ions, suggesting that the blocking site environment $\it{can}$ provide an additional 30 kcal/mol  to stabilize the Sr$^{2+}$ ion. 

\subsection{Conclusions}
Strontium (Sr$^{2+}$) and barium (Ba$^{2+}$) ions are comparable in ion size and have identical +2 charge, but Ba$^{2+}$ blocks bacterial potassium channels at the S4 binding site and Sr$^{2+}$ does not.  Our results suggest that in water, both ions adopt a local hydration structure  consisting of 8-fold skewed cubic water clusters that mimic binding sites for permeant ions in potassium ion channels.  But the smaller Sr$^{2+}$ requires an additional $\sim$~30 kcal/mol to take the ion out of the hydration environment relative to Ba$^{2+}$. This additional free energy can be compensated by the S4 binding site, due to a more favorable association free energy term balanced by an equally less favorable outer-shell free energy, leading to a transfer free energy of $\sim 0$. The transfer free energy for Ba$^{2+}$ is also $\sim 0$ and shows the same trends in contributions from the association and outer-shell terms. From the point of view of transfer free energies, these results suggest that both divalent ions should behave similarly and readily transfer between water and the S4 binding site of potassium channels.

The interesting observation on equivalent transfer free energies leads to the conclusion that the reason that Ba$^{2+}$ blocks bacterial potassium channels in the S4 binding site, and Sr$^{2+}$ does not, requires exploration of  the free energies for exchange of water ligands for channel ligands. The additional solvation free energy necessary to lose the same number of water molecules may hinder ligand exchange for Sr$^{2+}$ compared to Ba$^{2+}$.  That conclusion is supported by the observation that both Sr$^{2+}$ and Ba$^{2+}$ block mammalian potassium channels, where channel structure located in the intracellular environment may promote ligand exchange. That hypothesis will be tested in future work. 

In summary, these results suggest that differences in Sr$^{2+}$ and Ba$^{2+}$ blocking behavior do not
arise due to differences in free energies of transfer from water to the binding site.  Instead, the free energies associated with exchange of water ligands for channel ligands may account for the differences, and will be the subject of future work.

\section*{Methods}
\subsection*{AIMD simulations}
	The VASP $ab~initio$ molecular dynamics (AIMD) software was used with the PBE generalized gradient approximation \cite{Perdew:1996de} of electron density and the projector augmented wave (PAW) method \cite{blochl} for core electrons. A constant volume (NVT) ensemble simulation was run for 30 ps. The first 20 ps of data were used for equilibration and the last 10 ps were used for analysis. The simulations were conducted at a temperature of $T=350\pm21$ K and ambient pressure, $p=1$ atm, to be consistent with earlier studies of Ba$^{2+}$ hydration.\cite{Chaudhari:2014wb} A cubic box of 12.417 \AA~dimension along each side, ~with 64 waters and a single Sr$^{2+}$ ion, was simulated in periodic boundary conditions. 
	
	The system size and length of simulation time
	were sufficient to achieve a converged radial distribution function, confirmed by a separate classical MD simulation of the Ba$^{2+}$ system simulated for 50 ns.\cite{Chaudhari:2014wb} 
	In that box volume, the water density matches the experimental density of liquid water at standard conditions of room temperature and pressure used in experimental structural studies. The boundaries contained a background charge to neutralize the overall charge of the system. 
	With the PAW method,  the valence electronic orbitals were expanded in plane waves with a high kinetic energy cut-off of 29.4 Ry (400 eV) and 10$^{-4}$ eV was assigned as the convergence criteria for the electronic structure self-consistent iterations.  A time step of 0.5 fs was used and a non-local van der Waals correlation function applied to account for dispersion corrections using the vdw-DF2 method.\cite{Klimes:2009ei} 
	 
While experiments and simulation studies treat different ionic concentrations, experiments also show that Sr$^{2+}$ hydration structure is independent of ionic concentrations.\cite{persson}

\subsection*{Selection of the S4 blocking site}
	The initial configuration of the S4 blocking site, composed of four threonine residues (Thr)$_{2*4}$, was obtained from the crystal structure data for the bacterial potassium ion channel from  {\it Streptomyces lividans}, KcsA (PDB ID: 1K4C).\cite{Zhou:2001vo} The K$^+$ present in the crystal structure, K$^+$(THR)$_{2*4}$, was replaced by either Ba$^{2+}$ or Sr$^{2+}$ ion. We selected the 1K4C structure because it has the highest resolution (2 \AA) available for any potassium ion channel. 
	
	The KcsA crystal structure with Ba$^{2+}$ at the S4 site is also available (PDB ID: 2ITD), but at lower resolution (2.5 \AA).\cite{lockless} 
	Comparison between the 2ITD crystal structure and the computed lowest energy Ba$^{2+}$(THR)$_{2*4}$ structure indicates that the Ba$^{2+}$-O distances in both structures are comparable (FIG.~5).

\begin{figure}
	\includegraphics[width=3in]{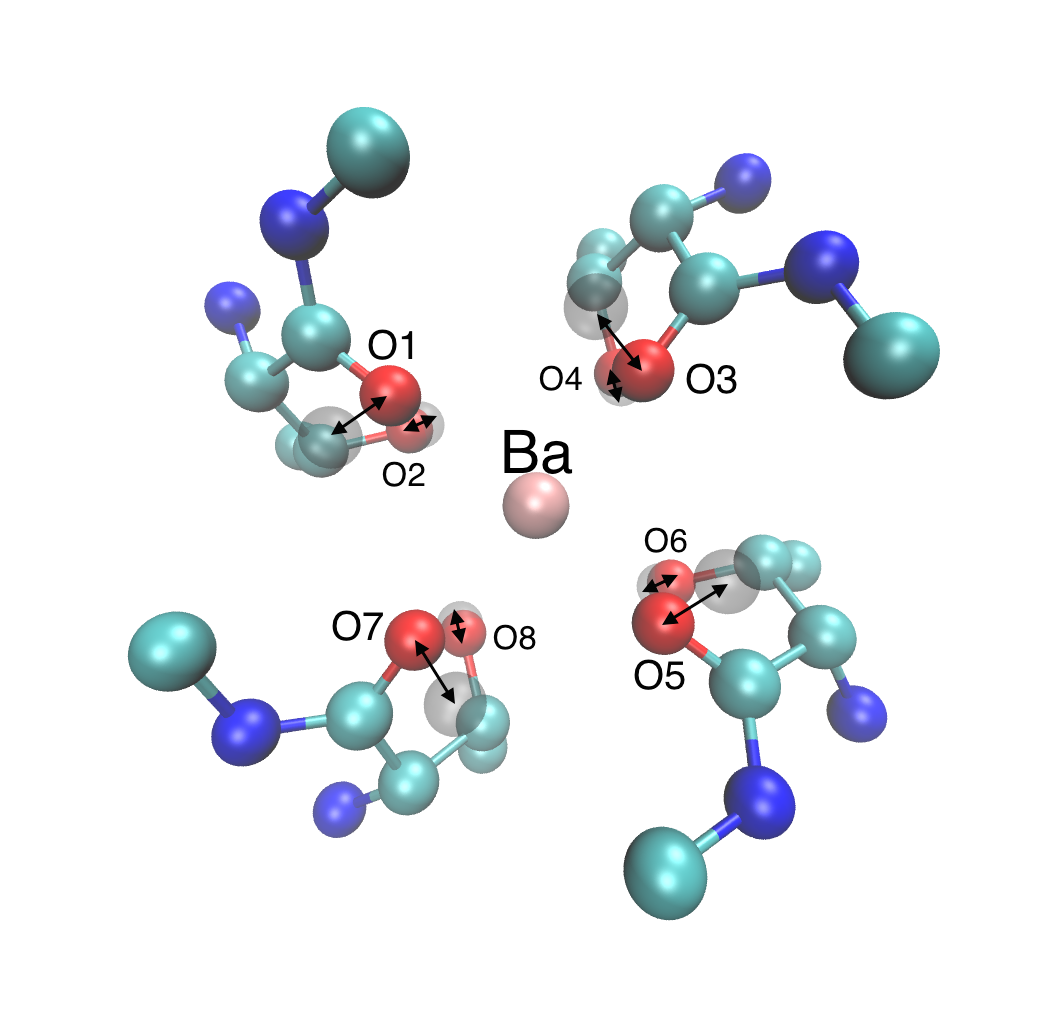}
	\includegraphics[width=3in]{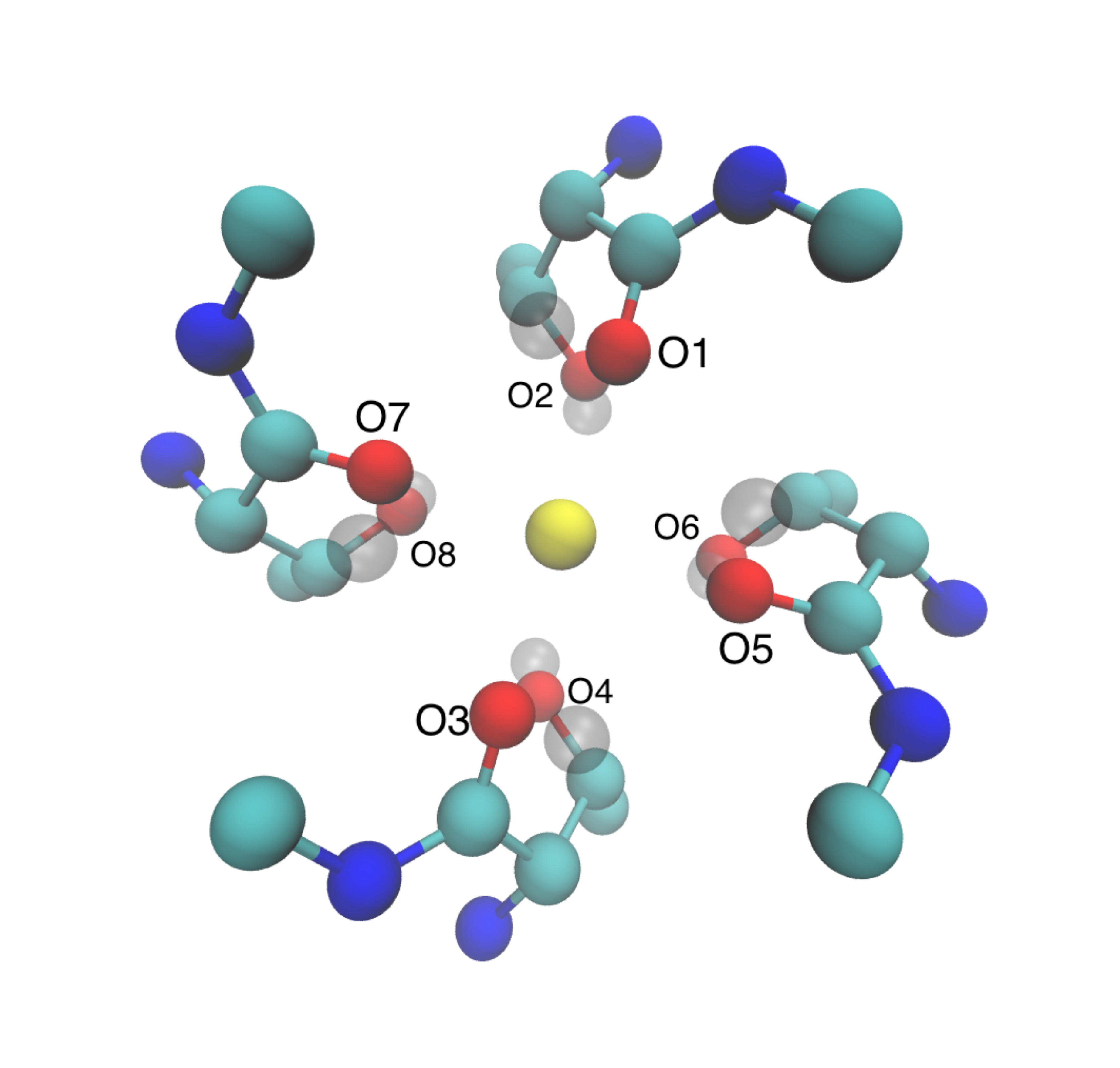}
	\caption{Pictorial representation of  oxygen atom displacements in the S4 site (THR$_{2*4}$) during electronic energy optimization. Oxygen atoms of crystal structure 1K4C (gray spheres) moved to new positions (red) after optimization
	    around (left) Ba$^{2+}$ (pink) and (right) Sr$^{2+}$ (yellow) ions.}\label{fig:fig2}
\end{figure}

\subsection*{Electronic structure calculations of the blocking site}
		 Electronic structure optimization and frequency calculations were done with the Gaussian suite of programs.\cite{g09} The TPSS functional and aug-cc-pvtz (O) and cc-pvdz (H) basis sets were used for  hydration studies of Ba$^{2+}$ and Sr$^{2+}$. A smaller basis set (6-31+G*) was used for C, H, N and O for binding site studies to decrease computational cost compared to the hydration calculations. 
		To be consistent, we recalculated hydration free energy $\muexBa$ and $\muexSr$ for ion hydration using the
		6-31+G* basis sets for H and O atoms of water (see FIG.~4 of the main text). Those  values are slightly different than the higher accuracy results for Ba$^{2+}$ \cite{Chaudhari:2014wb} and Sr$^{2+}$ hydration (FIG.~S3) due to the smaller basis sets used. Polarizable continuum model (PCM) was used to represent solvent behavior with various dielectric constant ($\varepsilon$) values. 

\section*{Acknowledgment}
We gratefully acknowledge support from the 
DTRA-JSTO CBD (IAA number DTRA10027IA-3167) and
Sandia's LDRD program.
Sandia National Laboratories (SNL) is a multimission laboratory managed and operated by National Technology and Engineering Solutions of Sandia LLC, a wholly owned subsidiary of Honeywell International Inc. for the U.S. Department of Energy's National Nuclear Security Administration under contract DE-NA0003525. This work was performed, in part, at the Center for Integrated Nanotechnologies (CINT), an Office of Science User Facility operated for the U.S. DOE’s Office of Science by Los Alamos National Laboratory (Contract DE-AC52-06NA25296) and SNL.


\end{document}